\documentclass[twocolumn, amssymb,aps,nopacs]{revtex4}
\usepackage{bm}
\usepackage{graphicx}
\usepackage{amsmath}
\usepackage{color}

\begin{document}

\title{Electrostatic shock dynamics in superthermal plasmas}

\author{S. Sultana\footnote{Email: basharminbu@gmail.com}}

\affiliation{Centre for Plasma Physics, Department of Physics and
Astronomy, Queen's University Belfast, BT7 1NN Northern Ireland,
UK}

\author{G. Sarri\footnote{Email: g.sarri@qub.ac.uk} } 

\affiliation{Centre for Plasma Physics, Department of Physics and
Astronomy, Queen's University Belfast, BT7 1NN Northern Ireland,
UK}

\author{I. Kourakis\footnote{Email: i.kourakis@qub.ac.uk; IoannisKourakisSci@gmail.com; www.kourakis.eu}}

\affiliation{Centre for Plasma Physics, Department of Physics and
Astronomy, Queen's University Belfast, BT7 1NN Northern Ireland,
UK}

\date{\today}
\begin{abstract}
The propagation of ion acoustic shocks in nonthermal plasmas is investigated, both analytically and numerically.
An unmagnetized collisionless electron-ion plasma is considered, featuring a superthermal (non-Maxwellian) electron distribution, which is modeled by a $\kappa$- (kappa) distribution function. Adopting a multiscale approach, it is shown that the dynamics of low-amplitude shocks is modeled by a hybrid Korteweg-de Vries -- Burgers (KdVB) equation, in which the nonlinear and dispersion coefficients are functions of the $\kappa$ parameter, while the dissipative coefficient is a linear function of the ion viscosity. All relevant shock parameters are shown to depend on $\kappa$: higher deviations from a pure Maxwellian behavior induce shocks which are narrower, faster and of larger amplitude.
The stability profile of the kink-shaped solutions of the KdVB equation against external perturbations is investigated
and the spatial profile of the shocks is found to depend upon
and the role of the interplay between dispersion and dissipation is elucidated.
\end{abstract}

\maketitle

\section{Introduction}

The study of shock waves in collisionless plasmas \cite{Forslund} has received, both from the experimental \cite{takeuchi1998,nakamura1999,nakamura2004,romagnani2008,heinrich2009,Kuramitsu} and theoretical \cite{zabusky1965,washimi1966,vladimirov1993,shukla2000,sahu2004,roy2008,masood2008} point of view, a great deal of attention in the last few decades. This widespread interest is justified by the central role that such structures play in many significant astrophysical scenarios such as cosmic rays generation \cite{McClements,Uchiyama} during supernova explosions \cite{Kulsrud}, strong fields generation in the bow-shock region \cite{Bale} and nonlinear dynamics of the solar wind \cite{Lee}, just to cite a few. Collisionless shock waves are routinely detected also in laser-plasma experiments which have therefore been proposed as smaller scale test-beds to systematically study astrophysical scenarios in the laboratory \cite{Zakharov}. The increasing level of detail offered by both laboratory measurements and space observations suggest the need  for continuous refining of the underlying theory. Most of the theoretical studies reported in the literature exploit the Korteweg-de Vries/Burgers (KdVB hereafter) equation which elegantly combines an interplay among nonlinearity, dispersion and dissipation effects in the generation and evolution of shock structures. However, the outstanding variety of plasma scenarios in which shocks may occur has made it difficult to develop a unified theoretical description of these intrinsically nonlinear excitations.

A major theoretical challenge in plasma modeling arises from the fact that most (e.g., astrophysical) plasmas may present significant deviation from a Maxwellian behavior, mostly due to the presence of accelerated electron populations \cite{vocks2003,gloeckler2006,vocks2008,malka1997,zhen2002}. Non-Maxwellian plasma particle distribution also occurs in the lab \cite{Hellberg2, Goldman, Sarri}, e.g., during high-intensity laser-matter interactions, where the generated plasma  clearly fails to thermalize within the short observation time interval, a fact reportedly affecting the measurable characteristics of nonlinear structures propagating in the plasma \cite{Sarri,Goldman}.
Fundamentally speaking, such energetic particles populate the superthermal region of the velocity distribution, and thus result in a long-tailed distribution function. A power-law behavior thus arises, at high velocities, a fact which motivated Vasyliunas \cite{vas1968} to
introduce a phenomenological long-tailed distribution, termed the kappa ($\kappa$) distribution (after the real valued $\kappa$ spectral parameter involved in it), which succeeds in reproducing the observed power-law dependence at high energies \cite{vas1968}, while remaining close to the Maxwellian curve at low (subthermal) velocities.
In the years that followed, the kappa distribution has been proven remarkably successful in fitting space observations (see, e.g., the various references in Refs. \onlinecite{pierrard, Livadiotis, Livadiotis2}) and, more recently, laboratory measurements \cite{Hellberg2,Sarri,Goldman}. Understandably, kappa theory has inspired a number  of theoretical studies \cite{summers1991,hellberg2009, sultana2010,baluku2010,AD}, with the ambition of elucidating the effect of superthermal populations on the plasma dynamics.

Despite its success on the space-observational test-bench, Vasyliunas' apparently ubiquitous distribution function remained in the grey area of phenomenology, lacking a rigorous foundation from first principle. In the meantime, alternative forms of the original kappa function have appeared \cite{Hau, hellberg2009}, citing all of which clearly goes beyond our scope here. The interested reader is referred to the interesting review and extensive discussion presented in Ref. \onlinecite{Livadiotis}.
A challenging non-Maxwellian-Boltzmannean paradigm appears to come from a different perspective, in Tsallis' non-extensive statistical-mechanical framework \cite{Tsallis1, Tsallis2}, a recently proposed generalization of Boltzmann statistics, which seems to incorporate a parametrized family of non-Maxwellian functions which correspond stationary equilibria which in fact maximize a generalized entropy, in analogy with the Boltzmann function and Gibbs entropy (smoothly recovered from the Tsallis theory at some limit). Not surprisingly, the Tsallis theoretical club has started to attract its own members within the plasma physics community \cite{Livadiotis, Lima2000, Bains}. As a natural consequence, the relation between kappa and Tsallis theories was investigated in Space plasmas \cite{Milovanov,Leubner2002} and, in fact, it was recently argued \cite{Livadiotis} on a rigorous basis that the very foundation of the kappa distribution lies in the Tsallis theory. Although this remains an open topic in the community, it would appear that the ubiquity of kappa distributions implies its deep foundations in Nature, which are still to be traced.

Our target in this article is to investigate, from first principle, the effect of excess superthermality on a fundamental problem in plasma physics, namely the propagation of electrostatic shocks evolving on the ionic scale.
Adopting the $\kappa$-theoretical framework in its original form \cite{vas1968,hellberg2009} and relying on the analytical toolbox presented in Ref.  \onlinecite{hellberg2009} (details of which are to be omitted), we shall develop a comprehensive formalism for one-dimensional ion-acoustic shock waves propagating in unmagnetized electron-ion plasma. Dissipation effects will be included by assuming non-negligible ion viscosity. The electron nonthermality is modeled by adopting a $\kappa$-distribution for the electrons. A nonlinear evolution equation for the electric potential will be derived, by using a multiscale (variable stretching) iterative scheme. The competing nonlinear and dispersive term(s) will be shown to be significantly dependent upon the $\kappa$ parameter which therefore significantly modifies the shock amplitude and width (compared to the Maxwellian case). Analytical and numerical techniques will be employed to solve the associated KdVB equation (discussion above),
and particular attention will be paid to the stability profile and the geometric characteristics of the shock excitations.

The layout of this article is as follows. The building blocks of the analytical model are presented in the following Section \ref{model} and its dispersion properties are analyzed, in Section \ref{lanalysis}. A multiscale perturbation technique is then employed in Section \ref{K-dVB}, leading to the working horse of the study, a Korteweg-de Vries/Burgers partial-differential equation (PDE) for the electrostatic potential. The shock-profile solutions of the K-dVB equation are presented and their dynamics is investigated in Section \ref{solution}. Our results are finally summarized in the concluding Section \ref{conc}.

\section{The Model \label{model}}

As mentioned in the introduction, we consider an unmagnetized electron-ion plasma where the electrons are assumed to be described by a $\kappa-$distribution function. The ions are assumed as a cold ($T_i = 0$) viscous population whereas the electrons will be considered as a fast neutralizing background. The electron inertia can in fact be neglected if we consider that ion-acoustic electrostatic waves move at phase velocity ($v_{ph}$) which is much higher than the ion thermal speed and yet in turn much lower than the electron thermal speed: $v_{th,i}\ll v_{ph}\ll v_{th,e}$. In a one-dimensional fluid description, the dynamics of the ions is governed by the following (normalized, dimensionless) set of fluid equations:
\begin{eqnarray}
\frac{\partial n_{i}}{\partial t}&+&\frac{\partial (n_{i} u_{i})}{\partial x} = 0\ ,  \label{ia1} \\
\frac{\partial u_{i}}{\partial t}&+& u_{i} \frac{\partial u_{i}} {\partial x}= -\frac{\partial \phi}{\partial x}+\eta \frac{\partial^{2}u_{i}}{\partial x^{2}}\ , \label{ia2} \\
\frac{\partial^{2}\phi}{\partial x^{2}}&=& n_{e} -  n_{i}\ . \label{ia3}
\end{eqnarray}
Here  $n_i$, $ u_i$, $\phi$ and $\eta$ represent the ion density (normalized to the equilibrium ion density), the ion  fluid velocity (normalized to the ion sound speed $c_{s}=( k_{B}T_{e}/m_{i})^{1/2}$), the electrostatic potential (normalized by $k_{B}T_{e}/e$)
and the ion kinematic viscosity, respectively.
Time and length are normalized in terms of the inverse of the ion plasma frequency ($\omega_{pi}^{-1}=[4\pi e^{2}n_{i0}/m_{i}]^{-1/2}$) and the plasma screening length ($\lambda_{D}=[k_{B}T_{e}/4\pi e^{2}n_{e0}]^{1/2}$) respectively. We note that $n_{e0}=n_{i0}$ was assumed to ascertain neutrality at equilibrium (the index `0' denotes the equilibrium quantities). Given this normalization, the ion viscosity $\eta_i$ is scaled into a (dimensionless) expression as $\eta = \eta_i / (\omega_{pi}\lambda_D^2)$. As usual, $m_i$ denotes the ion mass, $e$ is the electron charge, and $k_B$ is Boltzmann's constant; the ionic charge state was assumed to be $Z_i =1$. Ionic thermal effects are neglected, for simplicity.

The electron distribution, assumed to be non-Maxwellian, is modelled by a $\kappa-$type distribution \cite{hellberg2009, Livadiotis}; the electron density is thus given by:
\begin{equation}
\tilde n_{e}=n_{e0} \left[1-\frac{ e \Phi}{(\kappa-\frac{3}{2}) k_B T_e}\right]^{-\kappa +1/2}\ ,
\end{equation}
where dimensions are reinstated at this step (only) for the (physical) electric potential $\Phi$ and for the electron density $\tilde n_e$ (both latter quantities in physical dimensions, here, as opposed to the dimensionless counterparts $\phi$ and $n$).
The (reduced) electron density in Poisson's equation (\ref{ia3}) is thus taken to be of the form:
\begin{equation}
n_{e}=\biggl(1-\frac{ \phi}{\kappa-\frac{3}{2}}\biggr)^{-\kappa +1/2}\ , \label{ia4}
\end{equation}
as results from the first momentum of the one-dimensional $\kappa$-distribution function \cite{hellberg2009}. Here,
$T_e$ is a real constant, which corresponds to the electron temperature of the related Maxwellian plasma with the same particle and energy density \cite{hellberg2009}.  It is underlined that the $\kappa$ parameter holds a physical meaning if it belongs to the domain $(3/2,\infty)$; in the limit $\kappa\rightarrow\infty$, Eq. (\ref{ia4}) retrieves the Maxwellian expression \cite{hellberg2009}.

The power law dependence depicted in Eq. (\ref{ia4}) makes the differential equation (\ref{ia3}) of intractable analytically. In order to overcome this difficulty, we assume that any disturbance of the electric potential is small (i.e. $ \ll k_B T_e/e$). In this regime, a Taylor expansion of Eq (\ref{ia4}) around this parameter can be performed, allowing to express the electron density as $n_e \simeq 1 + c_1\phi + c_2 \phi^2 + {\cal{O}}(\phi^3)$.
The expansion parameters $c_1,c_2$ are only dependent upon $\kappa$ in the form:
\begin{equation}
c_1 = \frac{\kappa-1/2}{\kappa-3/2}, \hspace{1cm} c_2=c_1\frac{\kappa+1/2}{2(\kappa-3/2)} \label{c1c2}
\end{equation}
It is worth adding here that $c_{1}, c_{2} > 0$ $\forall$ $\kappa \in$ $(3/2,\infty)$.\\
Truncating the expansion at the second order, Eq. (\ref{ia3}) can be rewritten in the form:
\begin{equation}
\frac{\partial^{2}\phi}{\partial x^{2}}\approx 1-n_i +c_1\phi+c_2\phi^2 \label{phiapprox} \, .
\end{equation}
Since all the relevant quantities in the equations of interest now only refer to the ions, we will hereafter drop the subscript $i$ for simplicity of notation.

\section{Linear Dispersion Relation\label{lanalysis}}

A linear dispersion relation can be readily obtained by linearizing Eqs. (\ref{ia1}), (\ref{ia2}) and (\ref{phiapprox}), in the form:
\begin{equation}
\omega^{2}=\frac{k^{2}}{k^{2}+c_{1}} \ , \label{dr1}
\end{equation}
which reads, in physical units:
\begin{equation}
\omega^{2}=\frac{k^{2}\omega_{pi}^{2}}{k^{2}+(\sqrt{c_{1}}/\lambda_D)^2} \ . \label{dr2}
\end{equation}
\begin{figure}[!b]
\begin{center}
\includegraphics[width=0.38\textwidth]{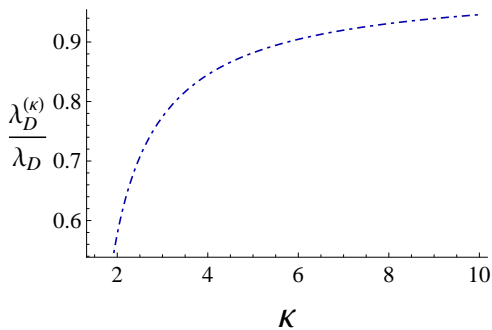}
\end{center}
\caption{(Color online) Dependence of the normalized $\kappa-$ dependent plasma screening length with $\kappa$.}
\label{fig.1}
\end{figure}

Eq. (\ref{dr2}) carries two fundamental consequences. The first is that $\omega$ is real for all $ k$, implying that no damped mode is present. This is due to the fact that a weak value of the damping coefficient was assumed, by scaling the ion viscosity as $\eta = \epsilon^{1/2}\eta_{0}$ (cf. the next Section), hence the latter does not appear in the dispersion relation.
Furthermore, the plasma screening length of a $\kappa$-distributed plasma is a function of $\kappa$: $\lambda_{D}^ {(\kappa)} = \sqrt{(\kappa-3/2)/(\kappa-1/2)}\lambda_D$, where $\lambda_D$ represents the screening length of a Maxwellian distributed plasma with same temperature and density (see Fig. \ref{fig.1}). The non-thermality of the plasma has thus the effect of \emph{reducing} the plasma screening length: the Maxwellian value is in fact retrieved in the limit case $\kappa \rightarrow\infty$.

It is worth noticing here that the dispersion relation obtained here coincides with Eq. (12) in Ref. \cite{sultana2010}, disregarding the magnetic field and setting $\nu = c_1$ therein.

In the long wavelength limit (i.e., for $k\ll 1$), Eq. (\ref{dr2}) reads
\begin{equation}
\frac{\omega}{k} \simeq \sqrt{1/c_{1}} \ c_{s} \simeq v_{ph}^{(\kappa)}  \ ,  \label{vph}
\end{equation}
where the $\kappa-$dependent phase speed is defined as $v_{ph}^{(\kappa)}=\sqrt{1/c_{1}} \ c_{s}$. The sound speed therefore deviates from the ion sound speed in the related Maxwellian plasma ($c_s$) by a factor $c_{1}^{-1/2}$. This implies that the sound speed (and subsequently the phase speed of an ion-acoustic wave) is reduced in the presence of superthermal electrons.

\section{Derivation of the K-dVB Equation \label{K-dVB}}

We proceed now to derive a $\kappa$-dependent evolution equation for propagating localized (constant profile) disturbances of the plasma dynamic  variables, by adopting a multiscale perturbation technique. This technique operates by stretching the spatial and temporal coordinate with the help of an infinitesimal parameter $\epsilon$ \cite{washimi1966,shukla2000}:
\begin{equation}
\xi=\epsilon^{1/2}(x-v_{ph}t) \ , \qquad \tau=\epsilon^{3/2}t \ , \label{ia10}
\end{equation}
where $v_{ph}$ is the ion excitation speed. The dependent variables $n$, $u$ and $\phi$ are perturbed around their equilibrium values in a power series in $\epsilon$ as:
\begin{eqnarray}
n&=&1+\epsilon n_{1}+\epsilon^{2} n_{2}+\epsilon^{3} n_{3} + \cdots \ , \nonumber\\
u&=&\epsilon u_{1}+\epsilon^{2} u_{2}+\epsilon^{3} u_{3} + \cdots \ , \nonumber\\
\phi&=&\epsilon \phi_{1}+\epsilon^{2} \phi_{2}+\epsilon^{3} \phi_{3} + \cdots \ . \label{ia15}
\end{eqnarray}
We consider a weak damping by scaling the ion kinematic viscosity as $\eta=\epsilon^{1/2}\eta_{0}$.\\
Substituting Eqs. (\ref{ia10}) and (\ref{ia15}) into Eqs. (\ref{ia1}), (\ref{ia2}) and (\ref{phiapprox}), and separating the lowest order of $\epsilon$ (i.e. terms proportional to $\epsilon^{3/2}$) the first order perturbations are obtained as: $n_1= \phi_1/v_{ph}^2$, $u_1=\phi_1/v_{ph}$ and $v_{ph}=\sqrt{1/c_1}$. The latter expression outlines the fact that, at first order, the ion-acoustic shock propagates at the $\kappa$-dependent sound speed (as defined in Eq. \ref{vph}).\\
Proceeding to the second order of perturbation (i.e. terms proportional to $\epsilon^{5/2}$) a KdVB equation is obtained in the form:
\begin{equation}
\frac{\partial \phi_{1}}{\partial \tau}+A\phi_{1}\frac{\partial \phi_{1}}{\partial \xi}+B\frac{\partial^{3} \phi_{1}}{\partial \xi^{3}}=C\frac{\partial^{2} \phi_{1}}{\partial \xi^{2}} \ , \label{KdVB}
\end{equation}
where the nonlinearity coefficient $A$, dispersion coefficient $B$ and dissipation coefficient $C$ are defined as:
\begin{equation}
\begin{cases}
A=\frac{2(\kappa-1)}{2\kappa-1}\sqrt{1+\frac{2}{2\kappa-3}}, \\
 B=\frac{1}{2}\bigr(1+\frac{2}{2\kappa-3}\bigr)^{-3/2}, \\
 C=\frac{\eta_{0}}{2}. \label{ABC}
 \end{cases}
\end{equation}
In the dissipationless case, i.e. in the limit $\eta_{0}=0$, one can get the expected KdV equation and the corresponding nonlinearity and dispersion coefficients as reported in Ref. \onlinecite{baluku2010} for superthermal plasmas, upon substitution with $\sigma=T_{i}/T_{e} =0$ therein (in account of our cold ion fluid hypothesis).
\begin{figure}[!t]
\begin{center}
\includegraphics[width=0.38\textwidth]{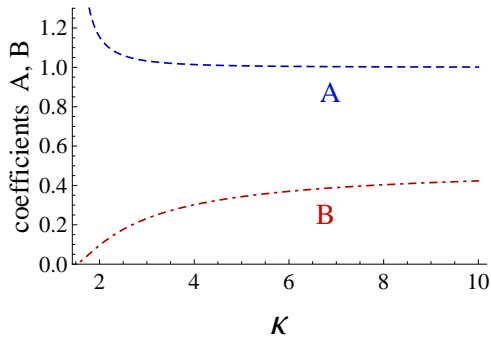}
\end{center}
\caption{(Color online) Dependence of the nonlinear coefficient $A$ and the dispersion coefficient $B$ upon the parameter $\kappa$.}
 \label{fig.2}
\end{figure}
It is clear from the above expressions that both the nonlinear term $A$ and the dispersion term $B$ are strongly influenced by $\kappa$. The nonlinear coefficient $A$ is higher for stronger superthermality (lower $\kappa$), while the dispersion coefficient $B$ is lower for stronger superthermality, as shown in Fig. \ref{fig.2}. The expected Maxwellian limit $A=1$ and $B=1/2$ is recovered for $\kappa\rightarrow\infty$; this is identical to, e.g.,  Eq. (28) in Ref. \cite{sghosh2002} (considering $\alpha=A, \ \beta=B, \ \delta=n_{i0}/n_{e0}=1$ and $\sigma=0$ therein).  It is worth noting again that both $A,B > 0$ $\forall \kappa$ $\in (3/2,\infty)$.

\section{Shock-profile solutions of the K-dVB equation \label{solution}}

The nature of the electrostatic shock-like solutions of the Eq. (\ref{KdVB}) strictly depends upon the relation between the equation coefficients $A,B,C$. In particular, given the overall weak dependence of $A$ upon the parameter $\kappa$ [see Eqs. (\ref{ABC}) and Fig. \ref{fig.2}], it is reasonable to assume that the behavior of the solutions of Eq. (\ref{KdVB}) will be mostly dictated by the interplay of the plasma dispersion and dissipation (depicted, in our model, by the parameters $B,C$). Moreover, given the exclusive relation of such parameters to $\kappa$ and to the ion viscosity $\eta$, it is reasonable to anticipate a certain relation between these two plasma quantities, in order to distinguish different regimes. In the following Section, we will therefore integrate Eq. (\ref{KdVB}), and study the stability of the obtained solutions, in different scenarios of arbitrary and strong dissipation.

\subsection{General case: an analytical solution}

We consider here the general case in which neither dissipation nor dispersion can be neglected in solving the KdVB equation shown in Eq. (\ref{KdVB}). First of all, in order to ease the derivation,
we will adopt a reference frame moving with the shock; this is translated in transforming the spatial and temporal coordinates into: $\zeta = \alpha (\xi-V\tau)$ and $\tau = \tau$ (the time dependence disappears, since we anticipate stationary profile solutions). Here $V$ represents the deviation of the shock speed from the plasma ion-acoustic speed and $\alpha^{-1}$ physically represents the shock width: qualitatively speaking, smaller values of $\alpha$ account for spatially extended shock profiles, and vice versa. With this transformation of coordinates, Eq. (\ref{KdVB}) reads:
\begin{equation}
-V\frac{d \phi_{1}}{d \zeta}+A\phi_{1}\frac{d \phi_{1}}{d \zeta}+B\alpha^{2}\frac{d^{3} \phi_{1}}{d \zeta^{3}}=C\alpha\frac{d^{2} \phi_{1}}{d \zeta^{2}} \ . \label{newcor}
\end{equation}
Integrating once and imposing boundary conditions of the form: $\lim_{\zeta\rightarrow\infty} \phi_1$ , $d\phi_1/d\zeta$ , $d^2\phi_1/d\zeta^2$ $=0$, Eq. (\ref{newcor}) reduces to:
\begin{equation}
B\alpha^2\frac{d^2\phi_1}{d\zeta^2}-C\alpha\frac{d\phi_1}{d\zeta}+\frac{A}{2}\phi_{1}^2-V\phi_{1}=0, \label{kdV}
\end{equation}
This equation is analytically solvable via the \emph{hyperbolic tangent} (tanh) approach \cite{malfliet1996,wazwaz2004,IKShSFV}. The general solution takes the form:
\begin{equation}
\phi_{1}(\xi, \tau)=D - \Phi_{0}\biggr[1+{\rm{tanh}}[(\xi-V \tau)/L_{0}]\biggr]^{2} \ , \label{phi0}
\end{equation}
where $L_0 = \alpha^{-1}=10B/C$ represents the shock spatial width and $D=\Phi_{max}=4\Phi_{0}=12C^{2}/(25AB)$ represents the shock amplitude (recall that $AB>0$ here). These values are obtainable by imposing that $V = 6C^2/(25B)$. This value of the velocity must be used in order to satisfy the boundary condition: $\lim_{\xi\rightarrow+\infty} \phi_1=V/A - 6C^2/(25AB)=0$ \cite{malfliet1996} \footnote{Admittedly, this appears as nothing but a mathematical artifact, that is, in general different asymptotic values may be obtained if one leaves $V$ arbitrary. However, in the process of  deriving the KdVB equation (\ref{KdVB}) in our context, it was assumed that $\phi = 0 $ in regions not yet visited by the shock; cf. the discussion in \cite{IKShSFV}.}. It is worth noting here that this solution corresponds to a shock travelling towards positive $\xi$ ($V>0$). It is readily seen that imposing opposite boundary conditions, a shock travelling in the opposite direction ($V<0$) would be obtained with the same absolute values of width, amplitude and velocity.

\begin{figure}[!b]
\begin{center}
\includegraphics[width=0.41\textwidth]{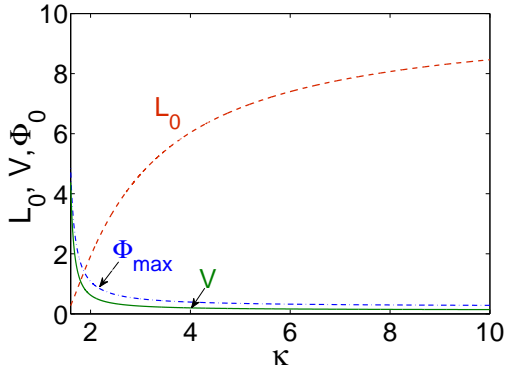}
\end{center}
\caption{(Color online) Dependence of the shock width $L_0$ (dashed red line), shock amplitude $\Phi_{max}=4\Phi_0$ (dot-dashed blue line) and shock velocity $V$ (solid green line) upon the non-thermality parameter $\kappa$ assuming $\eta_0 =1$.}
 \label{fig.3}
\end{figure}

A first interesting outcome of the analysis is that all the shock relevant parameters are functions of both $\kappa$ and $\eta_0$:
\begin{equation}
\begin{cases}
L_0 = \frac{10}{\eta_0}\left(1+\frac{2}{2\kappa-3}\right)^{-3/2}\\
\Phi_0 = \frac{3}{100}\eta_0^2\frac{(2\kappa-1)^2}{(\kappa-1)(2\kappa-3)}\\
V = \frac{3}{25}\eta_0^2\left(\frac{2\kappa-1}{2\kappa-3}\right)^{3/2} \label{LPV}
\end{cases}
\end{equation}
Recall that $V$ is the shock velocity increment above the acoustic speed (i.e., the shock velocity is $V_{shock}=c_{s}+V$), which is related to the viscosity $\eta_{0}$.
It is interesting to note that higher deviations from a pure Maxwellian behavior (i.e., smaller $\kappa$) implies shocks that are narrower, faster and with a more ample potential jump (see Fig. \ref{fig.3} where a reference value of the viscosity $\eta_0 =1$ has been considered). Nonetheless, for given $\kappa$, the product  $\Phi_{0} L_{0}^{2}$ is constant, thus retaining a well-known feature of soliton solutions of the KdV equation.

We now proceed to investigate the stability of a small perturbation ($\tilde{\phi_1}$) around the exact solution given in Eq.(\ref{phi0}) in the form: $\phi = \phi_1(\xi,\tau) + \epsilon\tilde{\phi_1}$, where $\epsilon$ is an infinitesimal parameter ($\epsilon\ll 1$).
Substituting $\phi$ in Eq. (\ref{kdV}) and linearizing with respect to $\tilde{\phi_1}$, we obtain a differential equation for the perturbation that reads:
\begin{equation}
B\alpha^2\frac{d^2\tilde{\phi_1}}{d\zeta^2}-C\alpha\frac{d\tilde{\phi_1}}{d\zeta}+\tilde{\phi_1} (A \phi_1-V) = 0\ . \label{perbeqn}
\end{equation}
This equation admits a general solution proportional to $e^{p\zeta}$, where the parameter $p$ obeys the characteristic polynomial:
\begin{equation}
\begin{cases}
B\alpha^2 p^2 - C\alpha p + A \phi_1-V = 0\\
p = \frac{C\pm\sqrt{C^{2}-4B(A \phi_1-V)}}{2B\alpha}\ , \label{polynomial}
\end{cases}
\end{equation}
Perturbations around the exact solution will therefore present an exponentially increasing (decreasing) behavior if $p$ is positive (negative) or an oscillatory behavior if $p$ is imaginary; in other words, the shock would sustain oscillatory perturbations if the discriminant of Eq. (\ref{polynomial}) is negative:
\begin{equation}
C^{2}-4B(A \phi_1-V)  < 0
\end{equation}
Substituting the expression for $\phi_1$ given in Eq. (\ref{phi0}), such a condition is translated into:
\begin{equation}
1+12[1+\tanh(\zeta)]^{2} < 0\ ,
\end{equation}
which is \emph{never} satisfied, regardless of the value of $\zeta$. The analysis therefore suggests that, in the general framework, oscillatory shock fronts are not allowed, regardless of the plasma characteristics.
\begin{figure}[!b]
\begin{center}
\includegraphics[width=0.43\textwidth]{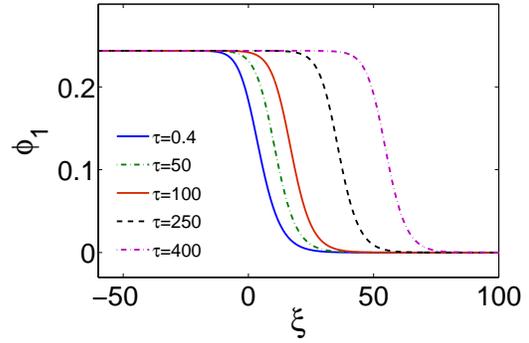}
\end{center}
\caption{(Color online) Temporal evolution of the ion-acoustic shock described by Eq. (\ref{phi0}) in a Maxwellian plasma ($\kappa\rightarrow\infty$). $C=\frac{\eta_0}{2}=0.5$ and $V=0.1$ have been considered.}
\label{fig.4}
\end{figure}
\begin{figure}[!b]
\begin{center}
\includegraphics[width=0.45\textwidth]{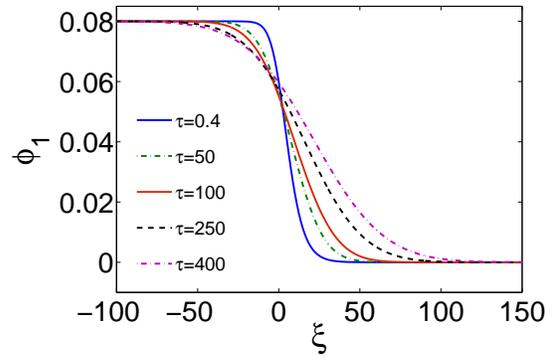}
\end{center}
\caption{(Color online) Temporal evolution of the ion-acoustic shock which is a solution for $C=\frac{\eta_0}{2}=2$ propagating in a superthermal plasma ($\kappa=3$), which is less dissipative ($C=\frac{\eta_0}{2}=0.2$).}
\label{fig.5}
\end{figure}

The two roots of Eq. (\ref{polynomial}) must be therefore real: $p_1,p_2 \in \mathbb{R}$. The formal solution of Eq. (\ref{perbeqn}) can thus  be written as:
\begin{equation}
\tilde\phi_{1} = F_{1} \ e^{p_{1}\zeta} + F_{2} \ e^{p_{2}\zeta}\ , \label{perturbsoln}
\end{equation}
where $F_1$ and $F_2$ are integration constants. Recalling Eq. (\ref{polynomial}), it is readily obtainable that the sum of the two roots $p_1,p_2$ must be equal to $C\alpha/(B\alpha^2)=10\eta_0>0$. The sum being positive, indicates that at least one of the roots must be positive implying that a small perturbation around the exact solution will be exponentially increasing in $\zeta$, therefore disrupting the shock. No perturbations of the shock profile can thus be supported in this regime. In other words, the solution (\ref{phi0}) provides a strictly monotonic profile (it is straightforward to show that the sign of the derivative of $\phi_1$ is uniquely defined) which is unstable to external perturbations.

A number of numerical simulations have been performed in order to verify our theoretical prediction. The propagation of the ion-acoustic shock has been analyzed for different plasma scenarios by numerical integration of the KdVB equation employing a Runge-Kutta 4 method. A time step of $10^{-4}$ and a spatial grid with size $0.1$ have been considered. As a test for the numerical code, the case of a shock propagating in a Maxwellian plasma ($\kappa\rightarrow\infty$) and moderate dissipation $\eta_0 =1$ has been first studied numerically. The shock is in fact seen to propagate unperturbed (see Fig. \ref{fig.4}).

\begin{figure}[!b]
\begin{center}
\includegraphics[width=7.5cm]{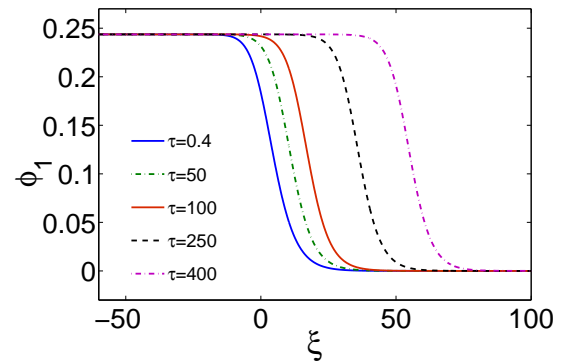}\hspace{0.2cm}

\huge{(a)}

\vspace{0.6cm}
\includegraphics[width=7.5cm]{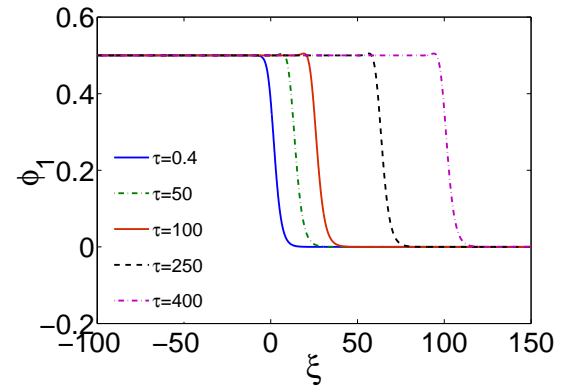}

\huge{(b)}
\end{center}
\caption{(Color online) (a) Temporal evolution of an ion-acoustic shock (using as initial condition the stable solution (\ref{phi0}) for a Maxwellian plasma), propagating in a strongly superthermal environment ($\kappa=3$). (b) Temporal evolution of an ion-acoustic shock (using as initial condition the solution which would be stable in a superthermal plasma, for $\kappa=3$), propagating in a Maxwellian environment. Here $C=\frac{\eta_0}{2}=0.5$ in both cases.}
 \label{fig.6}
\end{figure}
In order to trace the influence of the nonthermality (kappa) and damping ($\eta_0$) parameter(s), we may consider a propagating shock
crossing the interface between two plasmas with either different $\kappa$ or different viscosity $\eta_0$. In other words, we have considered the exact solution in plasma L, say, as initial condition to integrate the evolution equation (\ref{KdVB}) in plasma R (here L and R, stand for the left and right plasma regions, respectively, assumed to differ in the value of one parameter only, either $\kappa$ or  $\eta_0$, all other parameters being the same).

First, we have simulated the behavior of a shock crossing the interface between two plasmas with same $\kappa$ (here $\kappa=3$) but different viscosity (from $\eta_0 =4$ to $\eta_0=0.4$). The change in viscosity
only induces a widening of the shock width, clearly seen in Figure \ref{fig.5}, which is brought about by the lower dissipation encountered by the shock.

The situation is qualitatively different if we consider a transition in the superthermality of the plasma. If the shock penetrates  into a region of lower $\kappa$ (i.e. from a Maxwellian to a $\kappa=3$ plasma, in our simulation) it is seen to preserve its strict monotonicity and its width; see Fig. \ref{fig.6}(a). On the other hand, the reverse phenomenon (i.e. a shock which is stable in a superthermal plasma entering in a Maxwellian environment) presents a qualitatively different behavior. The shock in this case starts to develop a small hump in its spatial profile [see $\tau = 250, 400$ in Fig. \ref{fig.6}(b)]. These results can be interpreted in the following way. An increase in the $\kappa$ parameter (i.e. a smaller deviation from a pure Maxwellian behavior) implies an increase in the dispersive term $B$ in the KdV-B equation [see Eqs. (\ref{ABC})]. As it will be extensively discussed in the next section, oscillations at the front of an unstable shock occur only when the dispersion parameter dominates over the dissipative term. An abrupt increase of the dispersion parameter (as it is the case for an increase in the parameter $\kappa$) can verify this condition, thus inducing oscillations at the shock front. On the other hand, the reverse case (transition from a Maxellian to a $\kappa=3$ plasma) carries the only consequence of further decreasing the dispersion parameter, thus further forbidding the onset of oscillations at the shock front.

In the following, we proceed by adopting the same general treatment discussed in the previous section, yet considering the limit case of very weak ($C \gg B$) dispersion, compared to the damping coefficient $C$. The opposite case (i.e. dominant dispersion) will not be treated here, since it will lead to a natural convergence to the well-known KdV equation.

\subsection{Weakly dispersive/strongly non-Maxwellian limit \label{sdis}}

We now proceed by considering the \emph{particular} case in which the dispersion of the background plasma (modelled by the parameter $B$) is negligible. Given the dependence of $B$ upon $\kappa$ (Eq. (\ref{ABC}) and Fig. \ref{fig.2}), this is translated into studying the behavior of shock-like solutions of the KdV-B equation in the case of strongly non-Maxwellian background plasmas ($kappa$ in the vicinity of 3/2). In this particular regime, upon neglecting the dispersion coefficient $B$, Eq. (\ref{kdV}) becomes:
\begin{equation}
-V\frac{d \phi_{1}}{d \zeta}+A\phi_{1}\frac{d \phi_{1}}{d \zeta}=C\alpha\frac{d^{2} \phi_{1}}{d \zeta^{2}}, \label{nodisp}
\end{equation}
which admits a general solution of the form \cite{shukla2001}:
\begin{equation}
\phi_{1}(\xi,\tau)=\Phi_{1}\biggr(1-{\rm{tanh}}[(\xi-V \tau)/L_{1}]\biggr) \ . \label{nodispsoln}
\end{equation}
Eq. (\ref{nodispsoln}) embodies a shock structure with speed $V$, amplitude $\Phi_{1}=V/A$ and width $L_{1}=\alpha^{-1}=2C/V$.
In order to check the stability of such solution a similar analysis as the one performed in the previous section will be performed. A perturbed solution of the form $\phi = \phi_1(\xi,\tau) + \epsilon\tilde{\phi_1}$ will be again considered and substituted into Eq. (\ref{nodisp}) after having performed a spatial integration with the same boundary conditions discussed in the previous section. The linearization in $\tilde{\phi_1}$ leads to a differential equation for the perturbation of the form:
\begin{equation}
C\alpha\frac{d \tilde{\phi_1}}{d \zeta}-A\tilde{\phi_1} \phi_1+V\tilde{\phi_1}=0. \label{nodisp2}
\end{equation}
\begin{figure}[!b]
\centering
\includegraphics[width=6.5cm]{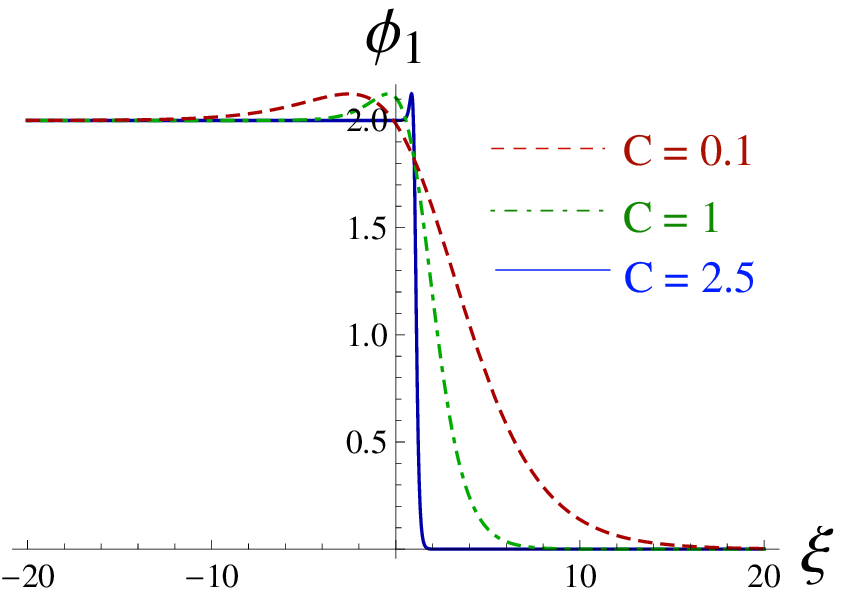}

\huge{(a)}

\hspace{0.6cm}
\includegraphics[width=6.5cm]{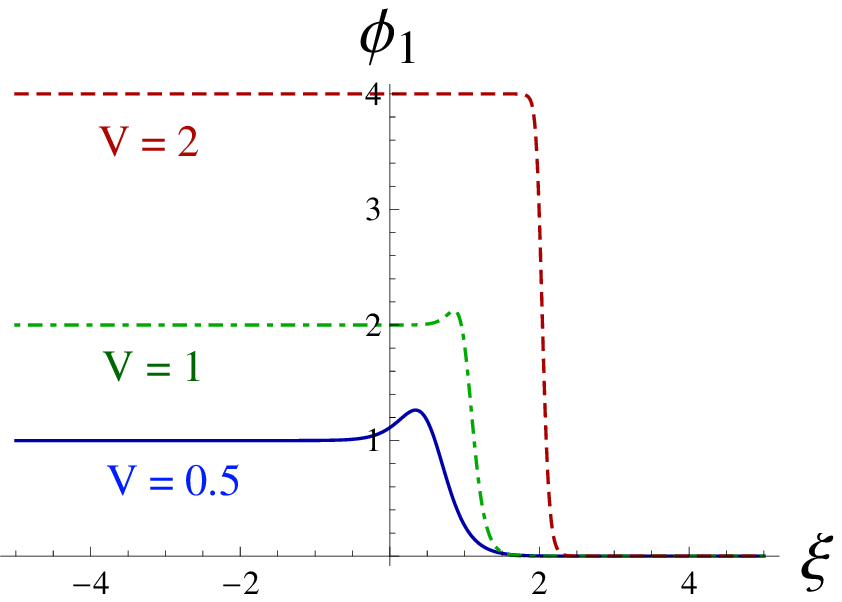}

\huge{(b)}
\caption{(Color online) (a) Shock profile for  different values of the dissipative term $C$, for a fixed velocity $V=1$ in the regime of negligible dispersion, i.e. relying on solution (\ref{nodispsoln}) with small perturbation; (b) Shock profile for  different values of the velocity $V$, for fixed dissipation  $C=0.1$ in regime of negligible dispersion. Here, time $\tau=1$ in both cases.} \label{fig.7}
\end{figure}
In this case, an analytical solution is obtainable by variable separation:
\begin{equation}
\tilde{\phi_1} = \Gamma \, \rm{sech}^{2}\zeta \ , \label{nodisp4}
\end{equation}
where $\Gamma$ represents the integration constant. Eq. (\ref{nodisp4}) depicts a KdV soliton-type solution, suggesting that the limit case of dispersionless plasmas supports shocks with infinitesimally perturbed fronts.
If we return to an external fixed reference frame (characterized by the spatial coordinate $\xi$), the width and amplitude of the perturbation will be dictated by the interplay of $C$ and $V$ (remember that $\zeta = (\xi-V\tau)/L_1$ with $L_1 = 2C/V$). In particular, higher propagation velocities (as much as smaller dissipation) will imply smaller perturbations. This conclusion is confirmed by numerical simulations of perturbed shock profiles shown in Fig. \ref{fig.7} for different values of ion viscosity [frame (a)] and propagation velocities [frame (b)].

It is interesting to test the behavior of shock solutions obtained in a purely dispersionless plasma, as they propagate through a region of not-negligible dispersion. Again, given the dependence of $B$ upon $\kappa$ (Eq. (\ref{ABC}) and Fig. \ref{fig.2}), this refers to a scenario in which a shock propagating in a highly superthermal zone ($\kappa$ near 3/2) enters a ``more Maxwellian'' (higher $\kappa$) region. Arguably, such a physical scenario reasonably also models the temporal evolution of a shock through a plasma which is undergoing a progressive thermalization, which is a situation often encountered in experiments. Analytically, this can be tested by substituting Eq. (\ref{nodispsoln}) into  Eq. (\ref{kdV}), taking into account that $\Phi_{1}=V/A$ and $L_{1}=2C/V$. This leads to an equation of the form:
\begin{equation}
-\frac{B V^{3}}{2A C^{2}} \, \rm{tanh}\zeta (\rm{tanh}^{2}\zeta - 1) = 0\ . \label{resteqn}
\end{equation}

This is equivalent to saying that a shock which is stable in the dispersion-less case, will preserve its stability in a dispersive plasma only if the condition (\ref{resteqn}) is satisfied. Although, strictly speaking, the latter condition is only satisfied (for arbitrary $\zeta$) if $B=0$, we may interpret the equality sign in the relation in a loose ($\simeq 0$) sense, thus allowing for different parameter combinations which may approximately satisfy this condition, in practical terms. The most trivial cases to consider are either $B=0$ or $C\rightarrow\infty$ (which both recover the dispersionless evolution of the system) or $\rm{tanh}\zeta = \pm 1$ (i.e. $\zeta\rightarrow\infty$ which suggests that a shock profile will keep to be stable in the regions far from the potential jump, regardless of the dispersion).
\begin{figure}[!b]
\begin{center}
\includegraphics[width=0.45\textwidth]{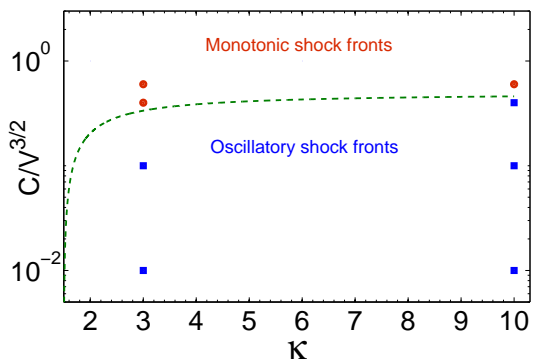}
\end{center}
\caption{(Color online) Threshold for the stability of dispersion-less shock fronts as a function of the non-thermality parameter $\kappa$. The points depict the initial conditions for the numerical simulations shown in the following.}
 \label{fig.8}
\end{figure}
However, a perturbation on the shock profile might be negligible also if $B V^{3}/2A C^{2} \ll 1$. Retrieving the dependence of $A,B$ upon the non-thermality coefficient $\kappa$, a condition of the following form is obtained:
\begin{equation}
\frac{C}{V^{3/2}}\gg \frac{\kappa-3/2}{\sqrt{2(\kappa-1)(2\kappa-1)}}\ . \label{critcondition}
\end{equation}
The above equation gives a $\kappa$-dependent condition which determines the nature (monotonic or oscillatory) of a dispersionless shock front in a dispersive plasma (cf. Figs. \ref{fig.9}-\ref{fig.12}).

\begin{figure}[!h]
\begin{center}
\includegraphics[width=6.5cm]{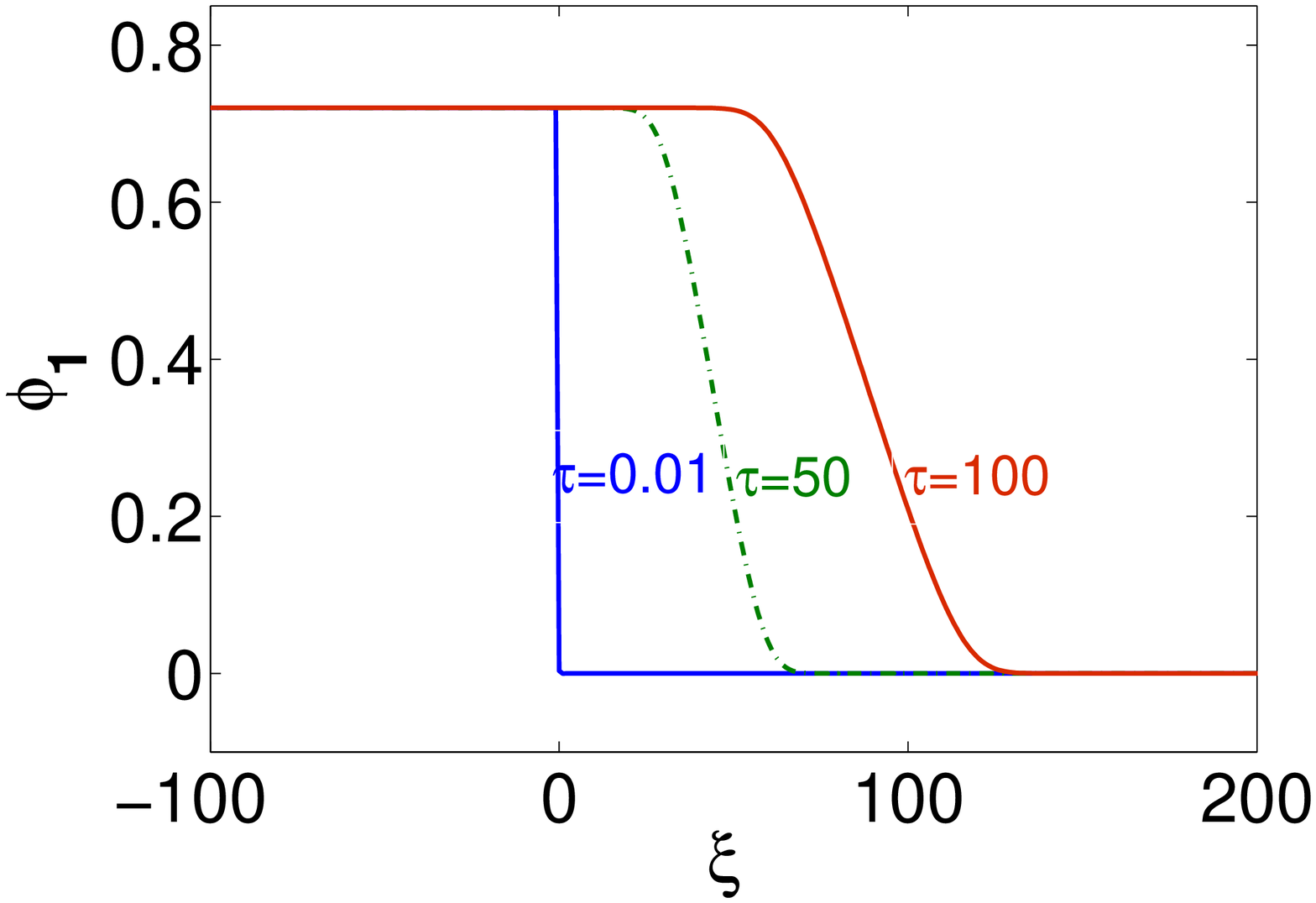}\vspace{0.3cm}
\includegraphics[width=6.5cm]{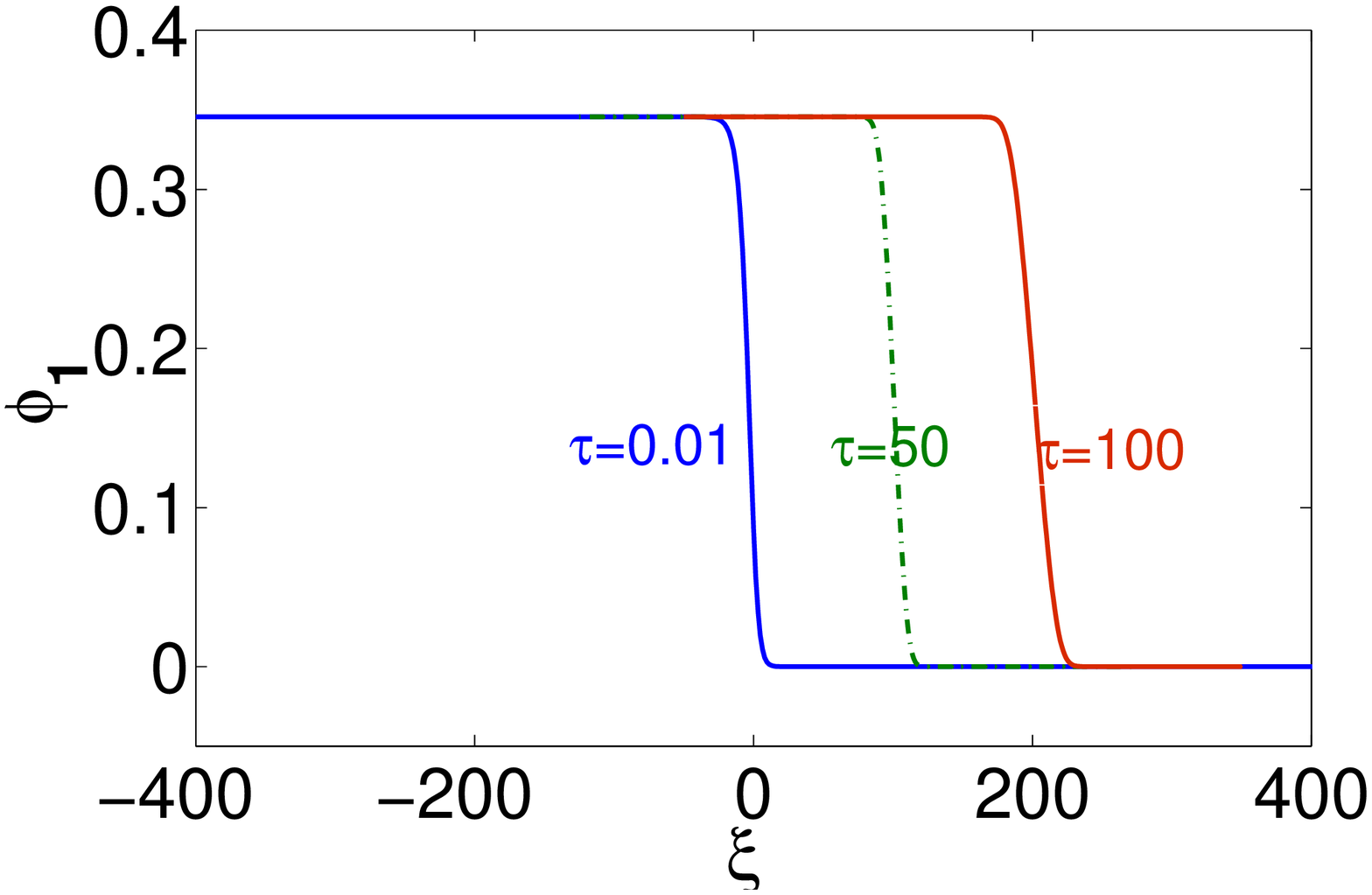}
\end{center}
\caption{(Color online) Temporal evolution of Eq. (\ref{kdV}) with initial condition given by Eq. (\ref{nodispsoln}) for a superthermal plasma ($\kappa=3$, upper panel) and a Maxwellian plasma ($\kappa\rightarrow\infty$, lower panel) for $C=\frac{\eta_0}{2}=0.6$. The shock profile exhibits a purely monotonic structure in both cases.}
 \label{fig.9}
\end{figure}

\begin{figure}[!h]
\begin{center}
\includegraphics[width=6.5cm]{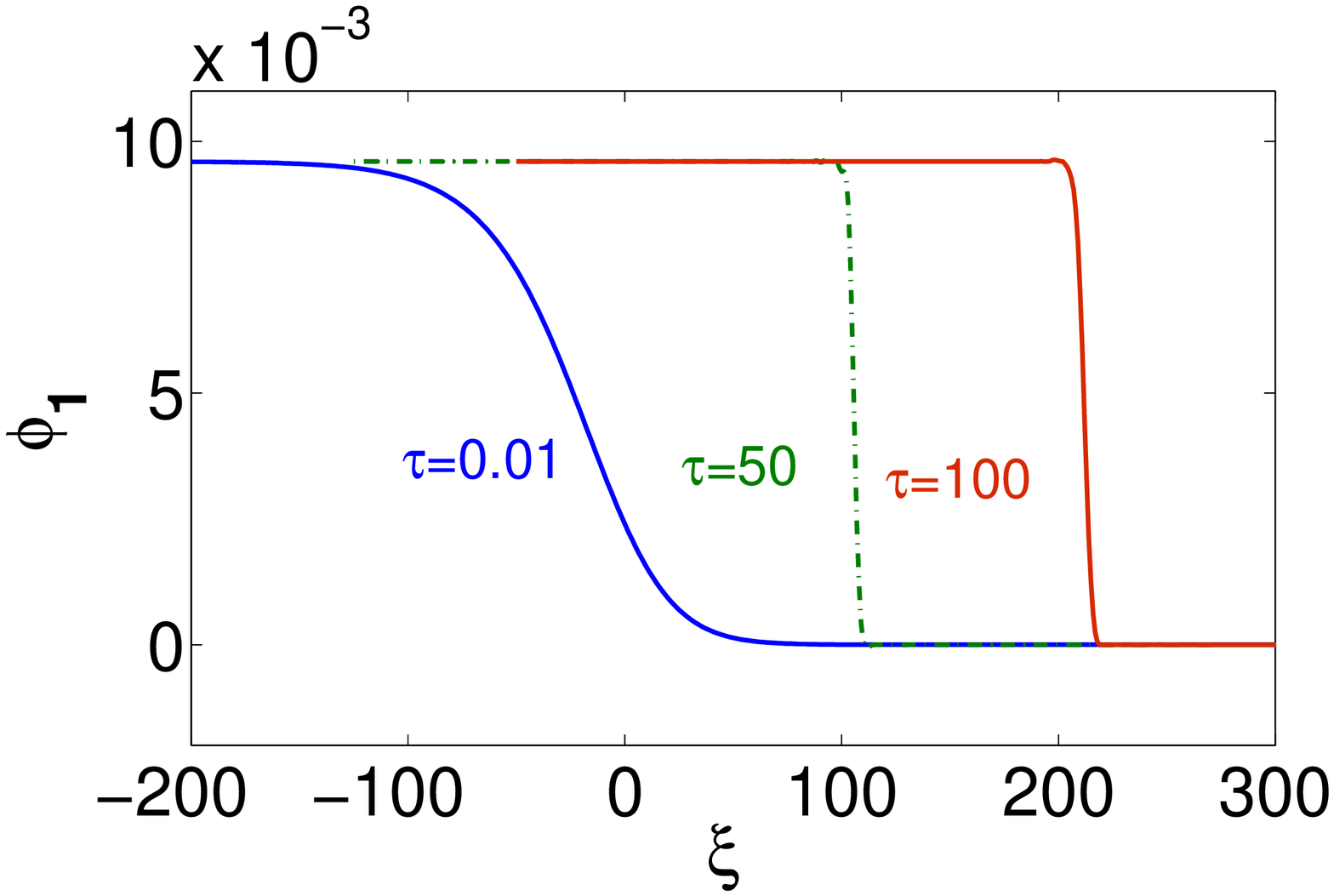}\vspace{0.3cm}
\includegraphics[width=7.3cm]{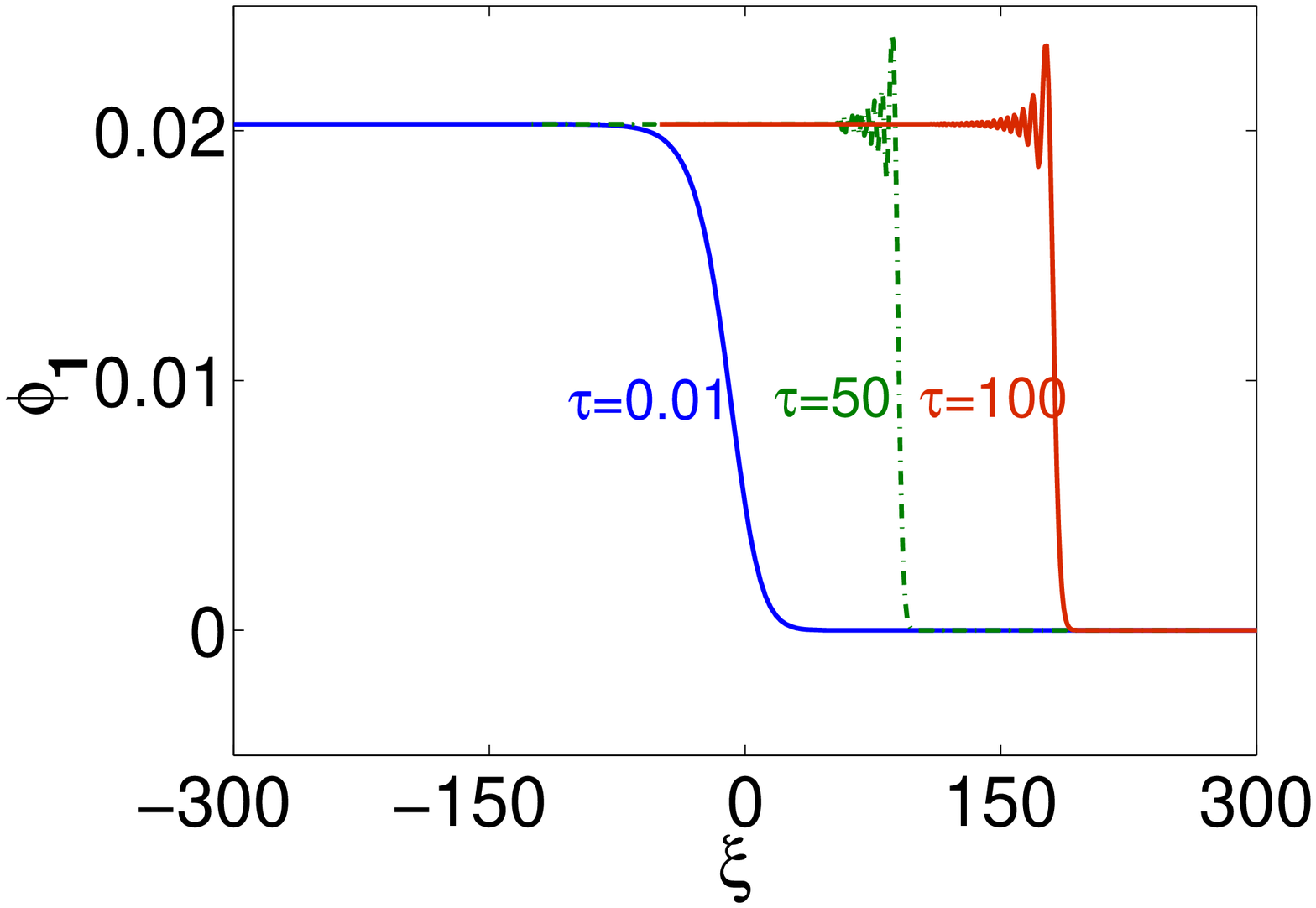}
\end{center}
\caption{(Color online) Temporal evolution of Eq. (\ref{kdV}) with initial condition given by Eq. (\ref{nodispsoln}) for a superthermal plasma ($\kappa=3$, upper panel) and a Maxwellian plasma ($\kappa\rightarrow\infty$, lower panel) for $C=\frac{\eta_0}{2}=0.1$. Oscillations are present in both cases.}
 \label{fig.10}
\end{figure}

In order to test the condition obtained above, a set of numerical simulations have been performed, and the results are shown in Figs. \ref{fig.9} - \ref{fig.12}.
The simulations, which were obtained in the same setup as described above, aimed at following the evolution in time of a shock profile which is stable in a dispersionless plasma [i.e. the exact solution of Eq. (\ref{nodisp}) depicted in Eq. (\ref{nodispsoln})] when it is forced to propagate through a plasma with non-negligible dispersion [whose KdVB equation is shown in Eq. (\ref{kdV})]. As mentioned above, one would expect the shock to preserve its monotonicity as long as the dissipation parameter is dominant compared to the dispersion one. This statement can be easily tested by noticing that, in our model, the dispersion and the dissipation coefficients are solely functions of $\kappa$ and of the ion viscosity parameter respectively [whose direct consequence is the threshold condition depicted in Eq. (\ref{critcondition})]. Different values of the viscosity $\eta_0$ have been therefore tested (see the red points in Fig. \ref{fig.8}) in both approximately Maxwellian ($\kappa = 10$) and strongly superthermal plasmas ($\kappa = 3$) in order to lay above and below the critical condition for oscillatory shocks. In all the simulations, a reference value of the velocity $V=1$ has been assumed. For $C=0.6$ (Fig. \ref{fig.9}) the shock is seen to be always monotonic, consistent with the analytical threshold which predicts the shocks being always monotonic regardless of $\kappa$. Similarly, for $C=0.1$ and $C=0.01$ (Figs. \ref{fig.10} and \ref{fig.11})
the shock always presents an oscillatory behavior, being below threshold for all values of $\kappa$. In both these cases it is interesting to note that the case $\kappa=3$ comprises less pronounced oscillations than the Maxwellian case, being closer to the threshold curve depicted in Fig. \ref{fig.8}. A peculiar situation can be found for $C=0.4$.
In this case the shock is expected to be monotonic for $\kappa=3$ (above threshold) and oscillatory for $\kappa=10$ (below threshold). Indeed this behavior is confirmed by the numerical results shown in Fig. \ref{fig.12}.

\begin{figure}[!h]
\begin{center}
\includegraphics[width=6.5cm]{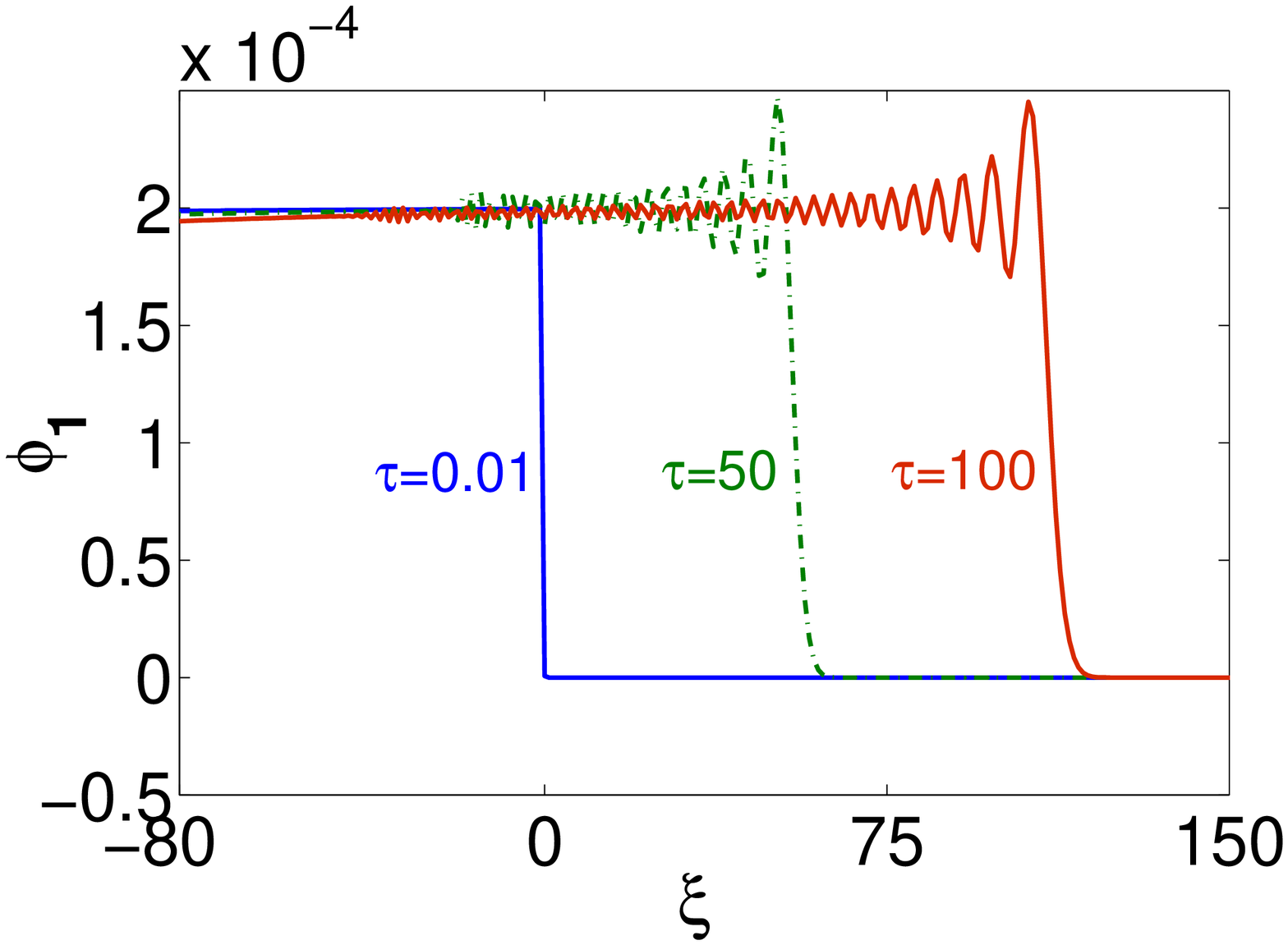}\vspace{0.3cm}
\includegraphics[width=6.5cm]{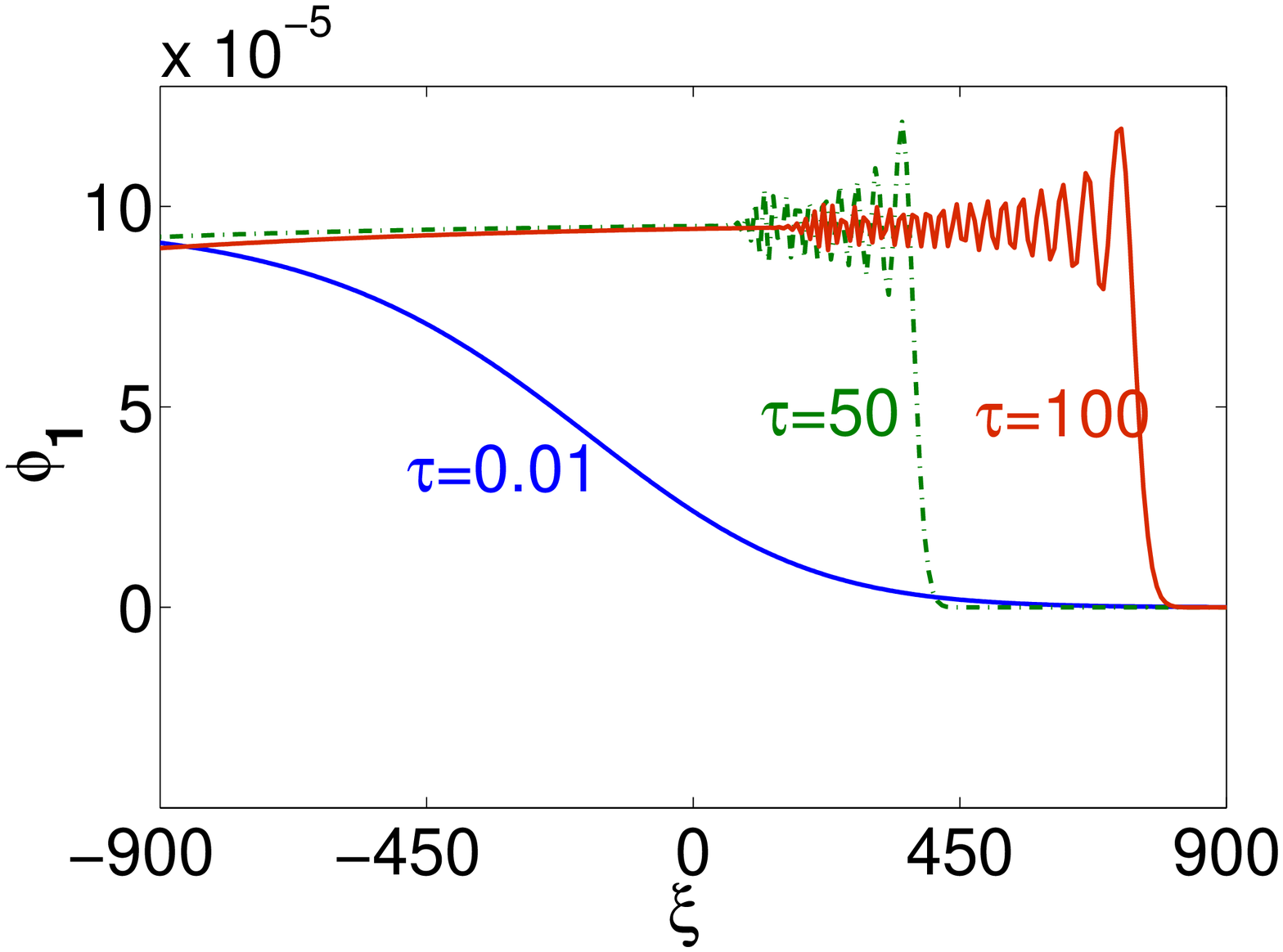}
\end{center}
\caption{(Color online) Temporal evolution of Eq. (\ref{kdV}) with initial condition given by Eq. (\ref{nodispsoln}) for a superthermal plasma ($\kappa=3$, upper panel) and a Maxwellian plasma ($\kappa\rightarrow\infty$, lower panel) for $C=\frac{\eta_0}{2}=0.01$. In both cases, the shock presents large oscillations.}
 \label{fig.11}
\end{figure}

\begin{figure}[!h]
\begin{center}
\includegraphics[width=6.5cm]{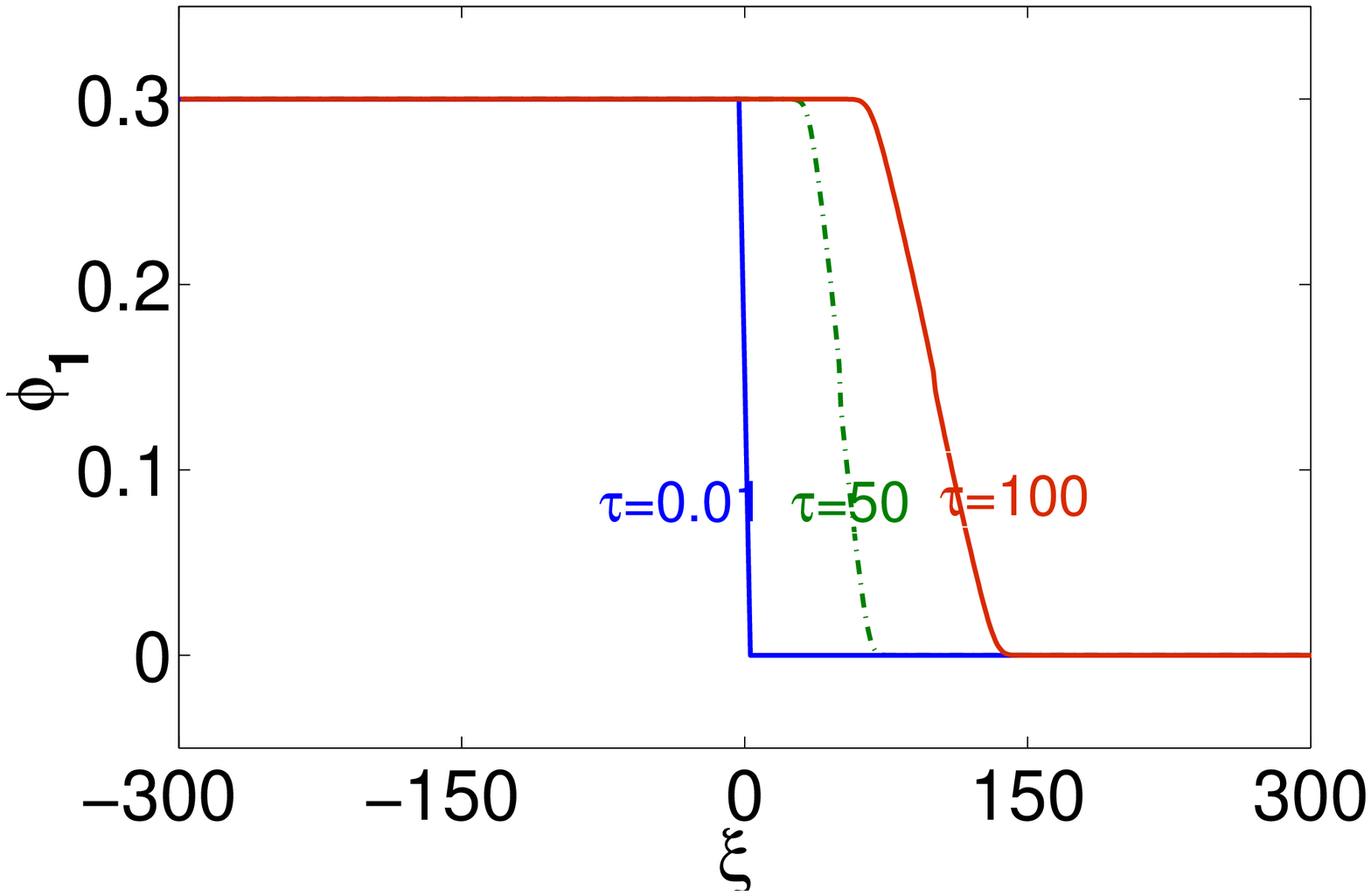}\vspace{0.3cm}
\includegraphics[width=7.3cm]{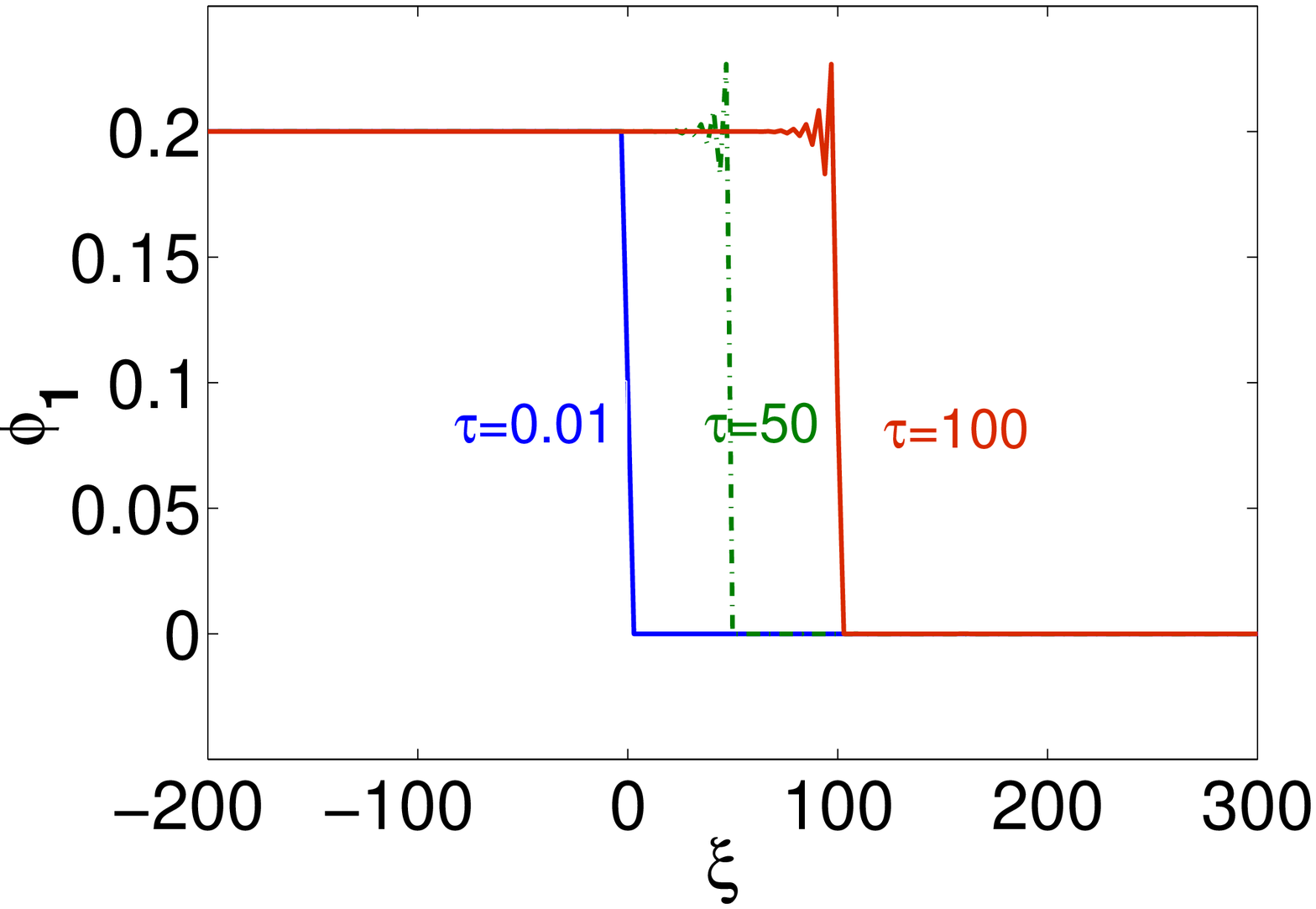}
\end{center}
\caption{(Color online) Temporal evolution of Eq. (\ref{kdV}) with initial condition given by Eq. (\ref{nodispsoln}) for a superthermal plasma ($\kappa=3$, upper panel) and a Maxwellian plasma ($\kappa\rightarrow\infty$, lower panel) for $C=\frac{\eta_0}{2}=0.4$. The shock presents oscillations only for the Maxwellian case.}
 \label{fig.12}
\end{figure}

\section{Conclusion \label{conc}}

 We have presented a theoretical study of the propagation dynamics of ion-acoustic shock waves in a plasma (in a one-dimensional geometry) characterized by an excess in the superthermal electron population, where the electrons obey the $\kappa-$distribution function, is presented. An unmagnetized electron-ion plasma has been assumed, with dissipation accounted for by considering the (constant) ion kinematic viscosity.

A Korteweg-de Vries-Burgers equation was derived for the electrostatic potential via a multiscale perturbation technique, and then analyzed analytically and numerically. The nonlinearity and dispersion coefficients are shown to be dependent upon the superthermality index $\kappa$, while the resulting damping coefficient is constant and only depends on the ion kinematic viscosity. Increasing the deviation from a Maxwellian behavior (i.e., considering smaller $\kappa$ values) simultaneously induces a decrease of the dispersion coefficient and an increase of the nonlinearity counterpart. As a consequence, ion-acoustic shocks in non-thermal plasmas are narrower, faster and of larger amplitude, as compared to the Maxwellian case.

Different types of shock solutions were analytically shown to exist depending on the relation between the dispersion and the dissipation coefficients. Although a strictly monotonic profile solution is analytically obtained in the general cases, substantial differences between the regimes of strong and weak dissipation have been found and confirmed by matching numerical integrations of the Korteweg-de Vries-Burgers equation. The transition of ion-acoustic shocks through the interface between plasmas of different characteristics (different non-thermality or viscosity) has been numerically studied, showing a sensible deformation of the shock profiles.

It appears appropriate here to add a comment to (formally) related work brought to our attention in the last stages of our study. Interestingly, a recent study has investigated superthermality effects on relativistic ion-acoustic shocks occurring in electron-positron-plasmas \cite{Pakzad1}, building upon a similar earlier investigation of Maxwellian plasmas \cite{Shah1}. We feel like paying due credit to Ref. \onlinecite{Pakzad1}, admitting that our algebraic setting is partly recovered from the model adopted therein, in the appropriate (classical, vanishing positrons) limit. We stress nonetheless the physical scope of our study at hand, which is distinct and presents no overlap with Ref. \onlinecite{Pakzad1}. On the other hand, we find that a number of methodological glitches (and consequent erroneous results) originating in Ref. \onlinecite{Shah1} have been inherited by Ref. \onlinecite{Pakzad1}, thus invalidating the (numerical, mainly) results therein. Therefore, not only the essential physics, but also the analytical formalism employed in those studies differ(s)  significantly from our results and method herein. We have chosen not to burden the presentation here with an extensive reference to these latter references, as the details will be discussed elsewhere \cite{IKShSFV}.
We note with interest that a later study by the same author \cite{Pakzad2011} was dedicated to the effect of the (Tsallis) $q-$nonextensive distribution on the dynamics of ion-acoustic shocks. The Tsallis distribution that was employed in that study \cite{Lima2000} is  \emph{not} identical, nor amenable to the one resulting from the original Vasyliunas \cite{vas1968} kappa theory that we employ here (refer to the discussion in the Introduction). Furthermore, we regret to find that the forementioned analytical discrepancy \cite{Shah1, Pakzad1,IKShSFV} arises in Ref. \onlinecite{Pakzad2011} as well, invalidating the results of an otherwise interestingly conceived study.
Concluding, neither the scope nor the details of our work present any overlap with the above studies.

Our simplified theoretical model represents a small yet steady step towards the rigorous understanding of the behavior of ion-acoustic shocks in non-thermal environments, which appear to be of fundamental importance in a wide range of astrophysical and laboratory scenarios. A better adherence to realistic scenarios of interest would require the relaxation of some assumptions adopted in the paper (such as taking into account of magnetic fields, multi-dimensionality and more realistic models for the ion viscosity) which will be subject of further work.

\acknowledgments
The authors are grateful to the UK Engineering and Physical Sciences Research Council (EPSRC) for financial support via grants No. EP/D06337X/1 and EP/I031766/1. G.S. acknowledges the Leverhulme Trust for financial support via grant ECF-2011-383. S.S. acknowledges a number of useful discussions with Dr. Bengt Elliason on the numerical part.

\appendix

\section{Derivation of the hybrid Korteweg -de Vries -- Burgers (KdVB) equation}

Although the method employed here may strictly speaking not be original, a brief outline of the perturbative methodology leading to the derivation of the KdVB equation (\ref{KdVB}).

Our starting point is the set of Eqs. (\ref{ia1}) - (\ref{ia3}). Anticipating a slow space, and a very slow time variation, according to the rationale underlying the KdV theory for acoustic electrostatic solitons \cite{washimi1966}, we introduce the coordinates $\xi=\epsilon^{1/2}(x-v_{ph}t) \ , \, \tau=\epsilon^{3/2}t$. One thus writes
 \begin{equation}
 \frac{\partial}{\partial x} = \epsilon^{1/2}\frac{\partial}{\partial\xi}\ , \qquad \frac{\partial}{\partial t} = -\epsilon^{1/2}v_{ph}\frac{\partial}{\partial\xi} + \epsilon^{3/2}\frac{\partial}{\partial\tau}\ . \label{stretch2}
 \end{equation}

 Applying the stretching (\ref{stretch2}) and the expansion (\ref{ia15}) into the model equations (\ref{ia1}) - (\ref{ia3}), and isolating the contributions in  $\epsilon^{3/2}$, one obtains a set of equations which can be solved to give the wave phase velocity
 \[
 v_{ph} = \sqrt{1/c_1} \,
 \]
where $c_1$ was defined in the text. Interestingly, the latter expression, here obtained as a compatibility condition, is exactly recovered via a different path if one considers the linear equations to obtain the ratio $\omega/k$ in the long wavelength limit; cf. the discussion in Section \ref{lanalysis}. (This simply confirms the well-known superacoustic nature of electrostatic acoustic solitons.) The first-order density and velocity perturbation(s) are also obtained at this order, in terms of the electrostatic potential (perturbation) $\phi_1$, respectively given by
 \[
n_1= \phi_1/v_{ph}^2\ , \qquad u_1=\phi_1/v_{ph}\ .
 \]
In the next higher order in $\epsilon$ ($\sim \epsilon^{5/2}$), we obtain the equations
\begin{eqnarray}
\frac{1}{v_{ph}^2}\frac{\partial \phi_{1}}{\partial \tau}&-&v_{ph}\frac{\partial n_{2}}{\partial \xi}+\frac{2}{v_{ph}^3}\phi_{1}\frac{\partial \phi_{1}}{\partial \xi}
+\frac{\partial u_{2}}{\partial \xi} = 0\ , \label{ia16}\\
\frac{1}{v_{ph}}\frac{\partial \phi_{1}}{\partial \tau}&-&v_{ph}\frac{\partial u_{2}}{\partial\xi}+\frac{1}{v_{ph}^2}\phi_{1}\frac{\partial \phi_{1}}{\partial \xi}
+\frac{\partial \phi_{2}}{\partial \xi}-\frac{\eta_{0}}{v_{ph}}\frac{\partial^{2} \phi_{1}}{\partial \xi^{2}} = 0\ , \label{ia17}\\
\frac{\partial^{2} \phi_{1}}{\partial \xi^{2}}&+& n_{2}-c_{1}\phi_{2}-c_{2}\phi_{1}^{2}=0 \ . \label{ia18}
\end{eqnarray}
Multiplying Eq. (\ref{ia16}) by $v_{ph}$ then summing with Eq. (\ref{ia17}) in order to eliminate
$u_{2}$,
one obtains
\begin{equation}
\frac{2}{v_{ph}} \frac{\partial\phi_1}{\partial\tau} - v_{ph}^{2} \frac{\partial n_{2}}{\partial\xi} + \frac{3}{v_{ph}^2} \phi_{1} \frac{\partial\phi_1}{\partial\xi} + \frac{\partial\phi_2}{\partial\xi} - \frac{\eta_{0}}{v_{ph}}\frac{\partial^{2}\phi_1}{\partial\xi^2} =0\, . \label{eliminu2}
\end{equation}
Now, differentiating Eq. (\ref{ia18}) with respect to $\xi$, one gets
\begin{equation}
\frac{\partial^{3}\phi_1}{\partial\xi^3} +\frac{\partial n_2}{d\xi} -c_{1}\frac{\partial\phi_2}{\partial\xi} - 2c_{2}\phi_{1}\frac{\partial\phi_1}{\partial\xi}=0\ . \label{eliminphi2}
\end{equation}

Multiplying Eq. (\ref{eliminphi2}) by $v_{ph}^2$, adding to Eq. (\ref{eliminu2}) and then multiplying by $v_{ph}/2$ leads us to write the resultant equation in the form of the Korteweg-de Vries Burgers (KdVB) equation given in (\ref{KdVB}).

\end{document}